\documentclass{aa}
\usepackage{graphicx}
\usepackage{amssymb}
\usepackage{longtable}
\usepackage{txfonts}

\newcommand{\Rstar}{\ensuremath{\mathrm{R}_{\star}}}
\newcommand{\rsun}{\ensuremath{\mathrm{R}_{\odot}}}
\newcommand{\mum}{$\mu$m}
\newcommand{\lam}{$\lambda$}

\newcommand{\kmsec}{\mbox{km~s$^{-1}$}}
\newcommand{\mdot}{\ensuremath{\dot{M}}}       
\newcommand{\msunyr}{M$_{\odot}$~yr$^{-1}$}

\newcommand{\teff}{\ensuremath{T_{\rm eff}}} 
\newcommand{\logg}{\ensuremath{\log g}}
\newcommand{\vinf}{\ensuremath{v_{\infty}}}

\newcommand{\ha}{H$\alpha$}
\newcommand{\hb}{H$\beta$}

\newcommand{\bra}{Br$\alpha$}

\newcommand{\brg}{Br$\gamma$}
\newcommand{\brten}{Br{\sc 10}}
\newcommand{\breleven}{Br{\sc 11}}

\newcommand{\pab}{Pa$\beta$}
\newcommand{\pag}{Pa$\gamma$}

\newcommand{\pfg}{Pf$\gamma$}
\newcommand{\pften}{Pf{\sc 10}}

\newcommand{\heii}{\ion{He}{ii}}
\newcommand{\hei}{\ion{He}{i}}

\newcommand{\ciii}{\ion{C}{iii}}
\newcommand{\oiii}{\ion{O}{iii}}
\newcommand{\civ}{\ion{C}{iv}}

\newcommand{\niii}{\ion{N}{iii}}
\newcommand{\feiii}{\ion{Fe}{iii}}

\newcommand{\siiv}{\ion{Si}{iv}}%
\begin{document}
\title{Modeling the near-infrared lines of O-type stars}
\titlerunning{Modeling the near-infrared lines of O-type stars}
\author{A. Lenorzer\inst{1} 
        \and M.R. Mokiem\inst{1} 
        \and A. de Koter\inst{1}
        \and J. Puls\inst{2}
       }
\authorrunning{A. Lenorzer et al.}
\offprints{A. Lenorzer (lenorzer@astro.uva.nl)} 
\institute{Astronomical Institute "Anton Pannekoek", 
           Kruislaan 403, NL-1098 SJ Amsterdam
      \and Institut f\"ur Astronomie und Astrophysik,
           Universit\"atssternwarte, Scheinerstr. 1, D-81679 M\"unchen 
          }
\date{Received 3 February 2004 / Accepted 30 March 2004}

\abstract{We use a grid of 30 line-blanketed unified stellar
photosphere and wind models for O-type stars; computed with the code
{\sc cmfgen} in order to evaluate its potential in the near-infrared
spectral domain. The grid includes dwarfs, giants and supergiants. We
analyse the equivalent width behaviour of the 20 strongest lines of
hydrogen and helium in spectral windows that can be observed using
ground-based instrumentation and compare the results with
observations. Our main findings are that: {\em i)} \hei/\heii\ line
ratios in the J, H and K bands correlate well with the optical ratio
employed in spectral classification, and can therefore be used to
determine the spectral type; {\em ii)} in supergiant stars the
transition from the stellar photosphere to the wind follows a
shallower density gradient than the standard approach followed in our
models, which can be mimicked by adopting a lower gravity in our
prescription of the density stratification.
{\em iii)} the \brg\ line poses a number of peculiar problems
which partly might be related to wind clumping,
and {\em iv)} the \bra\ line is an excellent mass-loss indicator. For
the first and last item we provide quantitative calibrations.

}

\maketitle
\section{Introduction}

A large fraction of the galactic population of massive stars lies
hidden behind tens of magnitudes of visual extinction. The reasons for
this are that massive stars are so rare that their typical distances
are a sizable fraction of the Galactic scale and that they are
concentrated in the Galactic disk. Therefore, these stars suffer from
obscuration by intervening molecular gas and dust clouds in the
line-of-sight.  Moreover, due to the short lifetimes of high-mass
stars, they are located in star forming environments. As they
typically form in the densest parts of these giant molecular clouds,
they spend a significant fraction of their life embedded in this natal
environment, before either moving out or breaking out.

Over the last decade several tens of massive stars were discovered in
the near-infrared spectral window, where extinction by intervening
dust is strongly reduced compared to the optical and ultraviolet
window (Hanson et al. 2002, Kaper et al. 2002, Kendall et al. 2003).
Studying the physical properties of these stars from their
near-infrared radiation alone is essential (for instance in relation
to their formation mechanism), but not an easy task. The J, H, K, and
L spectral windows contain relatively few lines, mostly of hydrogen
and helium. These lines are difficult to model as, for O-type stars,
most near-infrared lines are formed in the transition region from the
stellar photosphere (where the optical absorption spectrum originates)
to the super-sonic stellar wind (see e.g. Kudritzki \& Puls
2000). This makes a treatment of the stellar wind an integral aspect
of quantitative spectroscopic studies of O-type stars in the
near-infrared, perhaps excluding late-type O\,V stars of which the
stellar winds are relatively weak ($\mdot \leq 10^{-7}$ \msunyr). A
fundamental problem is that so far we have only a poor knowledge of
the way in which the density structure in the transition region and
lower part of the wind (up to a few times the sonic velocity)
behaves. Though the basic driving mechanism of stellar winds has been
identified (e.g. Castor, Abbott, \& Klein 1975, Abbott 1982,
Pauldrach et al. 1986) a fully self-consistent numerical
implementation of 
radiative line-driving is, at present, not feasible 
for models which have been constructed for the objective 
of atmospheric analyses.
Moreover, the inevitable assumptions made in
describing this theory are anticipated to have severe effects
on the physics of the transition region. 
As an example, the neglect of line source-function gradients when
using the Sobolev approximation might lead to erroneous values for the
radiative acceleration just in this transition region, at least in the
case of thin winds (cf. Owocki \& Puls 1999).

One may identify two essentially complementary approaches for making
progress in the development of near-infrared diagnostics that allow
for a characterisation of the basic stellar and wind properties of
O-type stars.
The first is to establish the near-infrared spectroscopic
characteristics of O stars with known properties from studies at other
wavelengths. One may then try to correlate the behaviour of these
lines with their basic properties and see if one can retrieve the same
information. This requires high-quality near-infrared spectra of a
large sample of MK standard stars. So far, a good coverage is
available only for the 2.0-2.2 \mum\ range (see Hanson et al. 1996,
2003). The second approach is to model the near-infrared lines using
state of the art techniques, and then to study the dependence between
spectral and basic properties.

So far modelling of the near-infrared spectral region has been done
for extreme early-type stars, i.e., Luminous Blue Variables (Najarro
et al. 1997), Of/WN stars (Crowther et al. 1995) and galactic centre
objects (Najarro et al. 1994). This study aims at an improved
treatment of normal O stars. With the use of sophisticated models that
include a detailed treatment of stellar winds and that properly
describe the spectrum, we study the predictions for the near-infrared
regime. We investigate to what extent the near-infrared lines can be
used to determine the spectral type, luminosity class, and mass loss
of O-type stars, focussing on lines that are observable using
ground-based instrumentation.

This paper is organized as follows: in Sect.~\ref{sec_grid} we
introduce our grid of models. Predicted equivalent widths (EW) of
near-infrared lines are presented and their dependence on model
parameters are discussed in Sect.~\ref{sec_trend}. Model prediction
are compared with observations gathered from the literature in
Sect.~\ref{sec_obs}. In Sect.~\ref{sec_param}, we present
near-infrared spectral classification schemes for O-type stars and a
means to determine wind properties. We end with conclusions.

\section{The grid of models}
\label{sec_grid}

\begin{table*}[htp!]
\caption{Stellar and wind parameters of our model grid.  For
supergiants, we list \logg\ values derived for both the spectroscopic
and evolutionary mass. Values of \teff\ and $R_\star$ for supergiants
correspond to models with evolutionary masses. Adopted abundances by
mass fraction are from Cox (2000): H\,=\,0.7023, He\,=\,0.2820, and,
in units of $10^{-3}$, C\,=\,3.050, N\,=\,1.100, O\,=\,9.540,
Si\,=\,0.699, Fe\,=\,1.360.}
  \begin{center}
  \begin{tabular}{cccccccccc}
  \multicolumn{10}{c}{\large Luminosity class V}\\
  \hline
  \hline
  Model & $T_{\rm eff}$ & ${M_\star}\over{M_\odot}$ & ${R_\star}\over{R_\odot}$ & $\log g$
  & $\log\left({L_\star}\over{L_\odot}\right)$ & $H_\star$ & $\log\dot{M}$ & $\beta$ 
  & $v_\infty$                                                              \\
      & (K) & & & (cm~s$^{-2}$) & & ($10^{-4}\,R_\star$) & ($M_\odot$/yr) & & (km~s$^{-1}$) \\
  \hline
  1V & 48618 & 68.9 & 12.3 & 4.094 & 5.882 & 8.308 & -5.375 & 0.8 & 3240 \\
  2V & 46070 & 56.6 & 11.4 & 4.075 & 5.722 & 8.375 & -5.599 & 0.8 & 3140 \\
  3V & 43511 & 45.2 & 10.7 & 4.032 & 5.568 & 8.971 & -5.805 & 0.8 & 2950 \\
  4V & 40964 & 37.7 & 10.0 & 4.012 & 5.404 & 9.046 & -6.072 & 0.8 & 2850 \\
  5V & 38406 & 30.8 & 9.3  & 3.987 & 5.229 & 9.314 & -6.369 & 0.8 & 2720 \\
  6V & 35861 & 25.4 & 8.8  & 3.951 & 5.062 & 9.712 & -6.674 & 0.8 & 2570 \\
  7V & 33306 & 21.2 & 8.3  & 3.924 & 4.883 & 9.934 & -7.038 & 0.8 & 2450 \\
  8V & 32028 & 19.3 & 8.0  & 3.915 & 4.783 & 9.988 & -7.252 & 0.8 & 2400 \\
  9V & 30820 & 17.7 & 7.8  & 3.897 & 4.697 & 10.16 & -7.445 & 0.8 & 2330 \\
 10V & 28270 & 14.8 & 7.4  & 3.864 & 4.502 & 10.40 & -7.913 & 0.8 & 2210 \\
  \hline
  \\
  \multicolumn{10}{c}{\large Luminosity class III}\\
  \hline
  \hline
  1III & 48132 & 82.8 & 15.1 & 3.996 & 6.042 & 9.098 & -5.125 &  1.0 & 3080 \\
  2III & 45363 & 68.4 & 15.0 & 3.919 & 5.934 & 10.04 & -5.239 &  1.0 & 2850 \\
  3III & 42595 & 56.6 & 14.8 & 3.848 & 5.813 & 10.82 & -5.397 &  1.0 & 2660 \\
  4III & 39815 & 47.4 & 14.7 & 3.777 & 5.690 & 11.53 & -5.585 &  1.0 & 2490 \\
  5III & 37049 & 39.0 & 14.7 & 3.692 & 5.564 & 12.63 & -5.791 &  1.0 & 2290 \\
  6III & 34282 & 32.6 & 14.7 & 3.614 & 5.430 & 13.50 & -6.051 &  1.0 & 2130 \\
  7III & 31504 & 27.4 & 14.7 & 3.539 & 5.283 & 14.20 & -6.377 &  1.0 & 1990 \\
  8III & 30127 & 25.1 & 14.8 & 3.495 & 5.211 & 14.69 & -6.551 &  1.0 & 1920 \\
  9III & 28781 & 23.5 & 14.8 & 3.464 & 5.134 & 14.73 & -6.758 &  1.0 & 1870 \\
 10III & 26011 & 20.2 & 15.0 & 3.386 & 4.970 & 15.18 & -7.230 &  1.0 & 1750 \\
  \hline
  \\ 
  \multicolumn{10}{c}{\large Luminosity class Ia}\\
  \hline
  \hline
  1Ia  & 47641 & 104.7\, 55.9& 18.6 & 3.916\, 3.639 & 6.206 & 9.502 & -4.896 & 1.0 & 3000 \\
  2Ia  & 44642 & 86.5\, 48.6& 19.6 & 3.788\, 3.533 & 6.139 & 11.63 & -4.922 & 1.0 & 2620 \\
  3Ia  & 41651 & 74.7\, 42.5& 20.6 & 3.681\, 3.431 & 6.062 & 12.95 & -5.014 & 1.0 & 2400 \\
  4Ia  & 38663 & 64.3\, 37.4& 21.8 & 3.566\, 3.327 & 5.981 & 14.44 & -5.134 & 1.0 & 2190 \\ 
  5Ia  & 35673 & 54.8\, 33.1& 23.1 & 3.446\, 3.223 & 5.892 & 16.15 & -5.295 & 1.0 & 1990 \\
  6Ia  & 32687 & 46.7\, 29.5& 24.6 & 3.322\, 3.120 & 5.795 & 17.90 & -5.509 & 1.0 & 1810 \\
  7Ia  & 31194 & 43.1\, 27.9& 25.4 & 3.260\, 3.068 & 5.741 & 18.70 & -5.644 & 1.0 & 1730 \\
  8Ia  & 29719 & 40.9\, 26.0& 26.2 & 3.211\, 3.012 & 5.683 & 18.67 & -5.815 & 1.0 & 1690 \\
  9Ia  & 26731 & 36.1\, 23.7& 28.1 & 3.094\, 2.909 & 5.563 & 19.35 & -6.202 & 1.0 & 1570 \\
 10Ia  & 23733 & 32.4\, 22.0& 30.5 & 2.976\, 2.805 & 5.427 & 19.53 & -5.409 & 1.0 & 740 \\
  \hline
\end{tabular}
\end{center}
\label{tab:StelPar}
\end{table*}

For this study we employ a grid of unified stellar photosphere and
wind models for O-type stars of luminosity class V, III and Ia. This
grid was constructed using the {\sc cmfgen} program of Hillier \&
Miller (1998), to which we refer for a full description. In short:
{\sc cmfgen} solves the equations of radiative transfer subject to the
constraints of statistical en radiative equilibrium, for an atmosphere
with an outflowing stellar wind. The ions included in the non-LTE
calculations are \ion{H}{i}, \ion{He}{i-ii}, \ion{C}{iii-iv},
\ion{N}{iii-v}, \ion{O}{iii-vi}, \ion{Si}{iv} and \ion{Fe}{iii-vii},
accounting for a total of approximately 20\,000 bound-bound
transitions. These reflect some 30\,000 lines and ensure a
self-consistent treatment of line blanketing, i.e., the cumulative
effect of the spectral lines, especially iron, on the stellar
atmosphere.

The grid consists of 30 models ranging in effective temperature,
$T_\mathrm{eff}$, from $\sim 24\,000 \; \mathrm{K}$ up to $\sim
49\,000 \; \mathrm{K}$, with 10 models for each luminosity class. The
stellar parameters are shown in Table~\ref{tab:StelPar}, with masses
derived from evolutionary tracks. For the basic stellar parameters we
employed the calibration from Vacca et al.(1996), which is based on a
set of plane parallel non-LTE H and He models that do not account for
line-blanketing. In the case of the last two models in the dwarf and
giant class and for the last three models in the supergiant class the
parameters were derived by extrapolating the relations found by these
authors for $T_\mathrm{eff}$, \logg\ and $M_V$.  For the stellar
radius we adopt the radius at which $\tau(R_\star) = 2/3$, with
$\tau(r)$ being the mean Rosseland optical depth corrected for
geometrical dilution (eq.\ 4 from Lucy 1976).
This does not exactly coincide with the radius give by Vacca et
al. However, the correction is minor and at most half a percent. To be
consistent we have included this correction in the values of \teff\
and \logg\ in Table~\ref{tab:StelPar}. For the supergiant models, we
also calculated a grid with gravities based on the (lower)
spectroscopic masses (see Vacca et al. 1996), in order to investigate
the dependence of the near-infrared lines on \logg\ (see e.g. Herrero
et al.\ 1992). For the chemical composition solar abundances from Cox
(2000) were incorporated, which are listed in the caption of
Table~\ref{tab:StelPar}.
Note in particular that our model grid comprises ``only'' models with
``normal'' Helium content, which has to be considered when comparing
our results with observations later on.
The reader should also note that the log g values of the
luminosity class V models in Table~\ref{tab:StelPar} do not reflect
ZAMS values. On the ZAMS there is a small negative correlation between
mass and \logg\ (e.g. Schaller et al. 1992). In the empirical
calibration from Vacca et al. this is reversed. However, for the major
objective and results of this study this has no influence.

The density structure in the photosphere is based on hydrostatic
equilibrium in an isothermal medium of temperature \teff. In that case
the density scale height is given by
\begin{equation}
  H = \frac{k T_\mathrm{eff}}{\mu m_\mathrm{amu} g_\mathrm{eff}}~,
  \label{eq:Hconstant}
\end{equation}
where $\mu$ is the mean molecular weight in atomic mass units ($m_{\rm
amu}$) and $g_\mathrm{eff}$ is the gravity at the stellar surface
corrected for radiation pressure by electron scattering. The density
structure near and beyond the sonic point is set by the velocity law
through the equation of mass-continuity. This velocity structure is
given by a standard $\beta$-law, which is smoothly connected to the
photosphere. The value of $\beta$ is set to 0.8 for the dwarfs and to
1.0 for the giant and super giants, as these stars have a tendency
toward higher $\beta$ (e.g. Groenewegen \& Lamers 1989; Puls et
al. 1996). The terminal wind velocity $v_\infty$ follows from a
scaling with the escape velocity $v_\mathrm{esc}$ (Abbott 1982, Lamers
et al. 1995). For stars with spectral type earlier than approximately
B2 this scaling implies that the ratio $v_\infty/v_\mathrm{esc}$ is
2.6. Stars with a later spectral type have a ratio of 1.3. This
discontinuity is referred to as the bi-stability jump and implies a
larger mass-loss rate and lower wind velocity for stars at the cool
side of this jump. This is the case for the coolest model in our grid,
Model\,10\,Ia. The mass-loss rates incorporated in the models are from
the theoretical predictions by Vink et al. (2000, 2001). These are
listed in Table~\ref{tab:StelPar} together with the terminal
velocities.

In the statistical equilibrium and radiative transfer calculation a
micro turbulent velocity of $v_\mathrm{turb} =
20\,\mathrm{km\,s^{-1}}$ was assumed for all lines. In the formal
solution of the radiative transfer equation, yielding the emergent
spectrum, we assumed micro turbulent velocities of 10 and 20
$\mathrm{km\,s^{-1}}$
(see e.g.\ Smith \& Howarth 1998; Villamariz \& Herrero 2000). Apart
from the broadening due to thermal motions and micro turbulence, Stark
broadening tables for \ion{H}, \ion{He}{i} and \ion{He}{ii} lines were
included.

As the effective temperature scale for O-type stars is currently being
revised using different models all accounting for line-blanketing
(e.g. de Koter et al. 1998; Martins et al. 2002; Repolust et
al. 2004), the spectral types attributed by Vacca et al. (1996) cannot
be applied to our models. Instead we use as a quantitative criterion
the ratio of the equivalent widths of the \hei\,\lam\,4471 and
\heii\,\lam\,4542 lines from Mathys (1988). This enables us to
unambiguously assign spectral types to our models.

\subsection{Limitations of the models}
\label{sec:limitations}

Though the models presented in this study are state-of-the-art, they
do (inevitably) contain a number of assumptions. With respect to the
prediction of the near-infrared spectrum the most important ones are:

\subsubsection{A constant photospheric scale height}

We assume a constant density scale height to describe the density
structure in the stellar photosphere (see Eq.~\ref{eq:Hconstant}). In
reality, the run of density in this regime follows from the equation
of hydrostatic equilibrium, i.e. it takes into account the exact
temperature structure,
the radiative pressure (from continua and lines) as well as changes in
the mean molecular weight. The lines that are expected to be affected
the most by this assumption are optical lines that are sensitive to
density, most notably the wings of Balmer lines. These lines are used
to derive the stellar gravity. Gravity determinations based on these
lines may therefore lead to a systematic overestimate of \logg\ of up
to 0.10 to 0.15 dex (P. Najarro, priv. communication). This effect is
less important for the strong near-infrared lines and especially for
supergiant models in which these lines are mainly formed beyond the
photosphere.

\subsubsection{The density structure in the transition zone}

A proper treatment of the density structure in the transition region
between the photosphere and the super-sonic wind requires solving the
equation of motion, taking into account all processes that are
responsible for the acceleration of the stellar outflow. Most
important, one should account for effects of radiation pressure on
spectral lines. This problem is at present too complex
to be solved for self-consistently in our models. Our approach is to
adopt a simple, empirical description of the density stratification in
this region, i.e.\ we smoothly connect the exponential increase of
velocity in the photosphere to a beta-type velocity law in the
supersonic regime.  This is achieved by taking
\begin{eqnarray}
   v(r) &=& \left( v_{\circ} + (\vinf - v_{\circ})
            (1 - R/r)^{\beta} \right) /  \nonumber \\
        & &  \left( 1 + (v_{\circ}/v_{\rm core}) \exp( [R-r] / H)
             \right)~,
   \label{eq:vlaw}
\end{eqnarray}
where $v_{\rm core}$ is the velocity at the inner boundary of the
model, and which, by means of the mass-continuity relation, sets the
density at $R_{\rm core}$. Typically, this density is chosen
sufficiently high to assure full thermalisation of the radiation
field. The velocity parameter $v_{\circ}$ is used to assure a smooth
transition from the photosphere to the supersonic wind. The latter is
prescribed by the terminal velocity \vinf\ and the parameter $\beta$,
essentially defining the steepness of the velocity law near and beyond
the sonic point.

An important outcome of this study is the ability to check whether
this commonly used representation of the density structure is able to
reproduce the properties of near-infrared lines, which are typically
formed in the transition region and/or lower part of the wind. The
major part of the discrepancies between observations and predictions
of these lines can likely be ascribed to shortcomings of the above
description.

\subsubsection{Clumping}

There is evidence that the stellar winds of early-type stars are
inhomogeneous on small length scales. Observational evidence exists
for Wolf-Rayet stars (Robert 1994, Hillier 1991) as well as for some
O-type stars (e.g. Eversberg et al. 1998, Bouret et al. 2003).
Theoretical indications for this effect are provided by Owocki et
al. (1988). One may anticipate that in clumped winds strong infrared
lines such as \bra\ and \brg\ and possibly \pfg\ and He~{\sc ii}\,(6-7)
will be affected the most. For recombination lines clumping effects
introduce a degeneracy in the quantity $\mdot / \sqrt{f}$, where $f$
is the clumping factor defined as $\overline{\rho} = f \rho$. Here it
is assumed that the inter-clump medium is void, and that
$\overline{\rho}$ is the unclumped wind density. Lines of varying
strength may be affected in different ways if the clumping factor
depends on radial distance. Potentially, the behaviour of the strong
near-infrared lines may yield constraints on the clumping
properties. As we want to focuss on other parameters in this first
study, we adopt an unclumped medium in our models.

\subsubsection{Turbulence}
In our models we assume a constant microturbulent velocity throughout
the photosphere and wind. On the basis of \bra\ and Pf$\alpha$
observations Zaal et al. (2001) found evidence for a gradient in the
turbulent velocity in the outer photospheres of late-O and early-B
dwarfs and giants. Such an increase in micro turbulence with radial
distance may also be present in O stars of earlier spectral type.

\subsection{Optical line trends}
\begin{figure}[!t]
\resizebox{8.5cm}{!}{\includegraphics{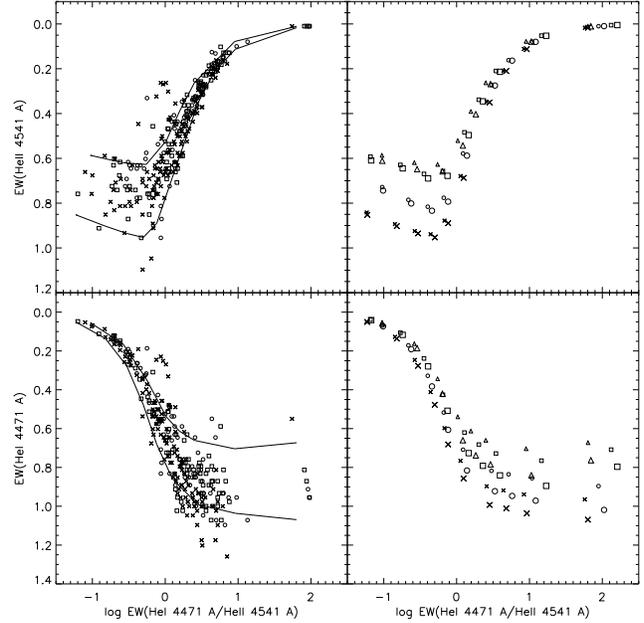}}
\caption[]{Left: Observed EW (in \AA) from Mathys (1988,1989), Conti
           \& Alschuler (1971), Conti \& Frost (1977), Bisiacchi et
           al. (1982) are plotted with crosses for dwarfs, circles for
           giants and squares for supergiants. Typical errors in the
           observations are 5 to 10 \%. The solid lines show the
           maximum and minimum values predicted in our models. Right:
           Equivalent width predictions for \hei\ are plotted for
           dwarfs (crosses), giants (circles), supergiants (squares)
           and lower surface gravity supergiants (triangles). Small
           symbols are for a turbulent velocity of 10 \kmsec, large
           symbols for turbulent velocity of 20 \kmsec.}
\label{optical}
\end{figure}

Before using our grid of models to investigate near-infrared lines, we
present their predictions for the two optical lines used for spectral
classification. Predicted EW are compared with observations gathered
from the literature and are shown in Fig.~\ref{optical}. The general
trends are well reproduced by the models, the largest deviation being
that the \heii\ line is slightly overestimated in dwarf models. This
discrepancy is also present in the grid of dwarf star {\sc cmfgen}
models presented by Martins et al.\ (2002). The grid of models covers
most of the span in equivalent width present in the observations. This
is not the case, however, when considering each luminosity class
separately. Indeed, stars with similar spectral type and luminosity
class show a range of stellar and wind parameters introducing a
scatter in the observed line EW. Furthermore variations in metallicity
and stellar rotation are not taken into account in our grid, but are
also a source of scatter. Still, the grid is suited for global studies
of basic parameters and is used for this purpose in the following
sections.

\section{Near-infrared line trends}
\label{sec_trend}
Before presenting our results, let us point out some principal problem
arising in (near) IR line formation (see also Mihalas 1978, Kudritzki
1979, Najarro et al.\ 1998, Zaal et al.\ 1999). Because $h\nu/kT \ll
1$ in the IR, stimulated emission becomes important, and deviations
from LTE in the line might become substantially amplified, compared to
the UV/optical case. Let $\delta = b_l/b_u -1$, with non-LTE departure
coefficients $b_l$ and $b_u$ for the lower and upper level of the
considered transition. By expansion, we then find for the IR line
source function (in units of the Planck-function)
\begin{equation}
\label{sline_ir}
\frac{S_{line}^{IR}}{B_\nu} \approx \frac{1}{1+\delta+\delta\frac{kT}{h\nu}},
\quad \frac{h\nu}{kT} \ll 1~,
\end{equation}
compared to the corresponding expression in the UV,
\begin{equation}
\frac{S_{line}^{UV}}{B_\nu} \approx \frac{1}{1+\delta}~, \quad \frac{h\nu}{kT} \gg 1~.
\end{equation}
In typical cases with $\lambda = 2 \mu m$ and $T=30\,000$~K, we thus
have
\[
\frac{S_{line}^{IR}}{B_\nu} \approx \frac{1}{1+5\delta}~.
\]
If now the upper level becomes overpopulated with respect to the lower
one (as it is typical for some IR transitions), $\delta$ is negative,
and the line source function can become much larger than the
Planck-function, leading to strong emission already for relatively low
wind densities. Even more important (and problematic), however, is the
fact that the response of the source function on small changes in the
population number ratio is much stronger than in the UV/optical
case. This means that small effects on the occupation numbers can have
substantial effects on the synthesized lines, and that apparently
large discrapancies between theory and observations may be caused by
relatively small inconsistencies. We will come back to this problem
when confronting our results for \brg\ with observations.

\begin{table}[t!]
\begin{center}
\begin{tabular} {lccr}
\hline
\hline
Line& Transition&Wavelength&Lines comprised in\\
    &           &\mum      &EW measurement \\  
\hline            
\hline
\pab     &3-5     &1.2822 &\heii\,(6-10)\,\lam1.2816\\
         &        &       &\hei\,(3p-5s)\,\lam1.2849\\
         &        &       &\hei\,(3d-5f)\,\lam1.2788\\
         &        &       &\ciii\,        \lam1.2794\\ 
\hline
\pag     &3-6	  &1.0941 &\heii\,(6-12)\,\lam1.0936\\
         &        &       &\hei\,(3d-6f)\,\lam1.0916\\
         &        &       &\ciii\,        \lam1.0920\\
\hline
\bra     &4-5	  &4.0523 &\heii\,(8-10)\,\lam4.0506\\
         &        &       &\hei\,(4f-5g)\,\lam4.0409\\
         &        &       &\hei\,(4f-5g)\,\lam4.0377\\
\hline
\brg     &4-7	  &2.1661 &\heii\,(8-14)\,\lam2.1652\\
\brten   &4-10	  &1.7367 &                         \\
Br\,11   &4-11    &1.6811 &                         \\ 
\pfg     &5-8 	  &3.7406 &                         \\
Pf\,9    &5-9     &3.2970 &                         \\
Pf\,10   &5-10    &3.0392 &                         \\
\hline
\hline
\heii    &5-7	  &1.1630 &                         \\
\hline
\heii    &6-7     &3.0917 &\heii\,(8-11)\,\lam3.0955\\  
         &        &       &\ciii\,        \lam3.0878\\           
         &        &       &\ciii\,        \lam3.0843\\           
         &        &       &\ciii\,        \lam3.0763\\           
\hline
\heii    &6-11	  &1.1676 &                         \\
\heii    &7-10	  &2.1891 &                         \\
\heii    &7-12	  &1.6923 &                         \\
\heii    &7-13    &1.5722 &                         \\
\hline
\hline
\hei     &2s-2p	  &2.0587 &                         \\
\hei     &3p-4s	  &2.1126 &                         \\
\hei     &3p-4s   &2.1138 &                         \\
\hei     &3p-4d	  &1.7007 &                         \\
\hei     &3p-5d   &1.1972 &                         \\
\hei     &3d-5f	  &1.2788 &                         \\ 
\hei     &4p-5d	  &3.7036 &                         \\
\hline
\hline
\civ     &3d-3p   &2.0802 &                         \\ 
\ciii    &7-8     &2.1151 &                         \\
\hline
\end{tabular}
\caption{Identification of the lines used in this study.}
\label{tab_ident}
\end{center}
\end{table}

    \begin{figure*}[!ht]
     \begin{center}
    \resizebox{\linewidth}{14cm}{\rotatebox{90}{\includegraphics{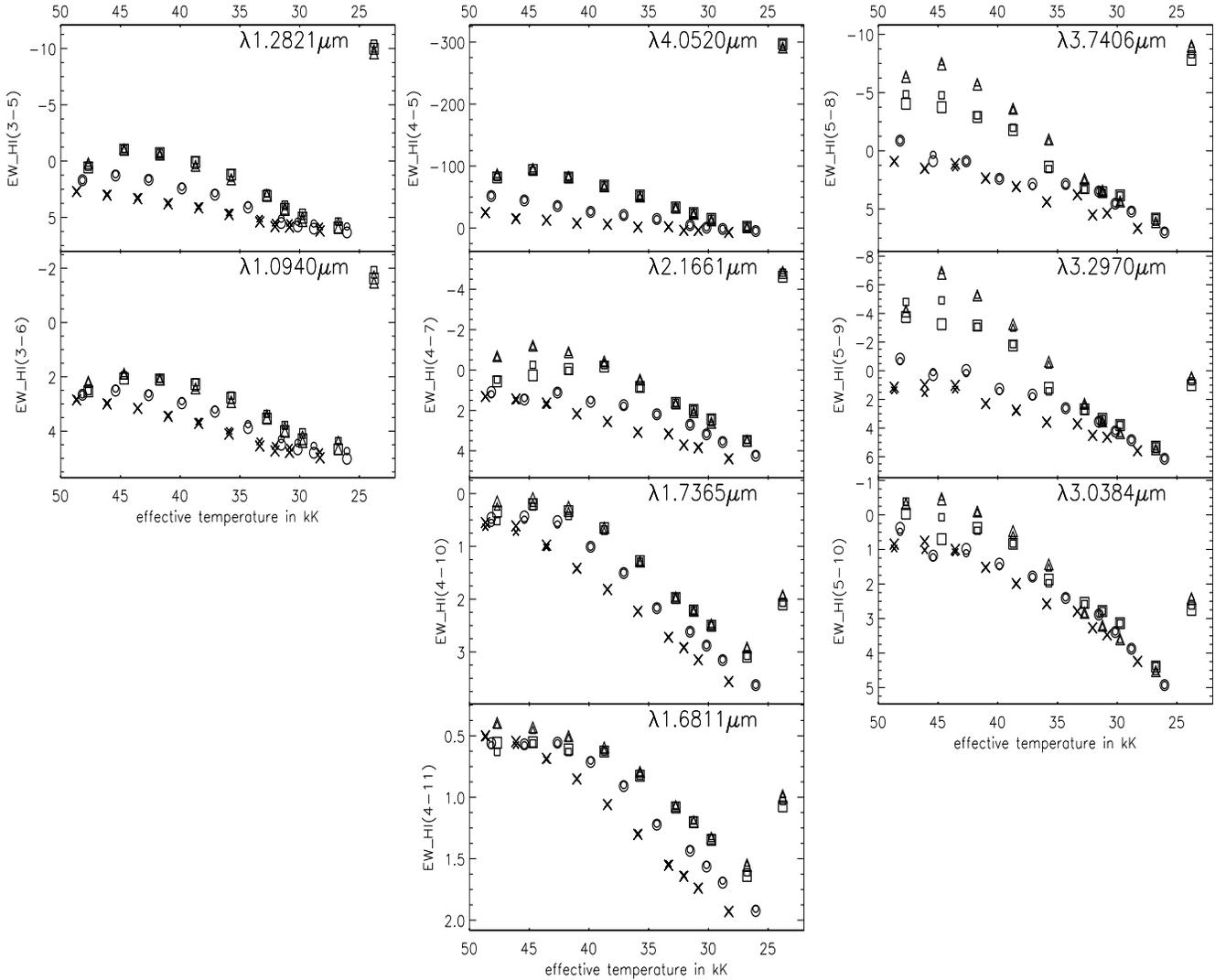}}}
    \caption{Equivalent width predictions (in \AA) for hydrogen lines
    are plotted for dwarfs (crosses), giants (circles), supergiants
    (squares) and lower surface gravity supergiants (triangles).
    Small symbols are for a turbulent velocity of 10 \kmsec, large
    symbols for a turbulent velocity of 20 \kmsec.}
    \label{fig_hyd}
    \end{center}
    \end{figure*}

\smallskip
\noindent
After this introductory remark, we will now present our equivalent
width (EW) predictions for the strongest hydrogen and helium lines in
the near infrared, and discuss their dependence on model
parameters. The lines are listed in Table~\ref{tab_ident}, together
with blends present in the wings of their profiles that are included
in the EW predictions. In the following a positive and negative EW
correspond, respectively, to an absorption line and an emission line.

\subsection{Hydrogen lines}
\label{sec_hyd}

The near-infrared domain includes several hydrogen lines from
different series: Paschen in the J band, Brackett in the H, K and L
band and Pfund in the L band. Higher members of the Pfund and
Humphreys series are also located in the K and L band, respectively.
A high density and a low hydrogen ionisation fraction are required to
make them observable. Although this is the case for cooler stars and
emission line stars, these lines are not reported in the spectra of
O-type stars and are not included in our model calculations which
include 13 levels of hydrogen.

We plotted EW predictions of the Paschen, Brackett and Pfund lines for
all models in Fig.~\ref{fig_hyd}. The \bra\ line is blended with the
\hei\,(4f-5g) and \heii\,(8-10) lines in all spectra, and with
\hei\,(4d-5f) for models with large wind velocities and mass-loss
rates. We defined the EW of \bra\ over the interval from 4.0 to 4.1
\mum, which includes these blends. \pab\ and \pag\ lines are
integrated over an interval that comprises the \hei\ lines (3d-5f) and
(3p-5s), and (3d-6f), respectively. The lines decrease in equivalent
width with temperature in the range 25 to 45 kK. This trend is steeper
for higher series, as well as for stronger lines within a series and
for lines in supergiants relative to other luminosity classes. The
lines show progressively lower equivalent widths for decreasing
luminosity class.

Two models do not follow this general trend: the hottest (1\,Ia) and
coolest (10\,Ia) supergiant models computed in this study. In
Model\,10\,Ia, the hydrogen lines have a more negative equivalent
width compared to neighbouring models whereas it is the opposite in
Model\,1\,Ia. This behaviour can be traced back to the adopted wind
parameters. Model\,10\,Ia is at the cool side of the bi-stability
jump, where it has a higher wind density (see Sect.~\ref{sec_grid}).
Model\,1\,Ia has a lower wind density than Model\,2\,Ia. The strongest
lines show the most pronounced emission, as their line forming region
extends further out in the wind. Differences with luminosity class are
smaller for photospheric lines (e.g. \breleven, \brten, \pften). Their
behaviour is mostly sensitive to temperature, and can therefore be
used to constrain the spectral type. This also holds for \pab, \pag,
\brg\ and \pfg, except for supergiants, which have large mass-loss
rates affecting also the $\beta$ and $\gamma$ lines.


The two grids of supergiant models show that the Pfund lines and \brg\
are sensitive to the surface gravity. As these lines form in the
region above the photosphere, but below the sonic point, they are
sensitive to changes in gravity. The strong \bra\ line is formed in a
much more extended region, i.e. also in the wind where the density
structure is set by the velocity law, and is less sensitive to gravity
effects. We have also investigated the effects of micro turbulence:
As only the micro turbulent velocity in the calculation of the formal
solution was modified, the occupation numbers are not influenced. Only
the line profiles are additionally broadened. For the lines
considered, the EWs differ only marginally between the two adopted
turbulent velocities (10 and 20 \kmsec).

The strength of most predicted near-infrared hydrogen lines show a
smooth behaviour as a function of model parameters. On top of this,
the Pfund lines, and to a lesser extent \brg, show small model to
model fluctuations. These may be intrinsic in nature (e.g. blends) or
the result of numerical effects (e.g.\ sampeling of the radius grid)
and do not significantly affect our results.

%

\subsection{\heii\ lines}
\label{sec_heii}

    \begin{figure*}[!t]
    \begin{center}
    \resizebox{\linewidth}{10.48cm}{\rotatebox{90}{\includegraphics{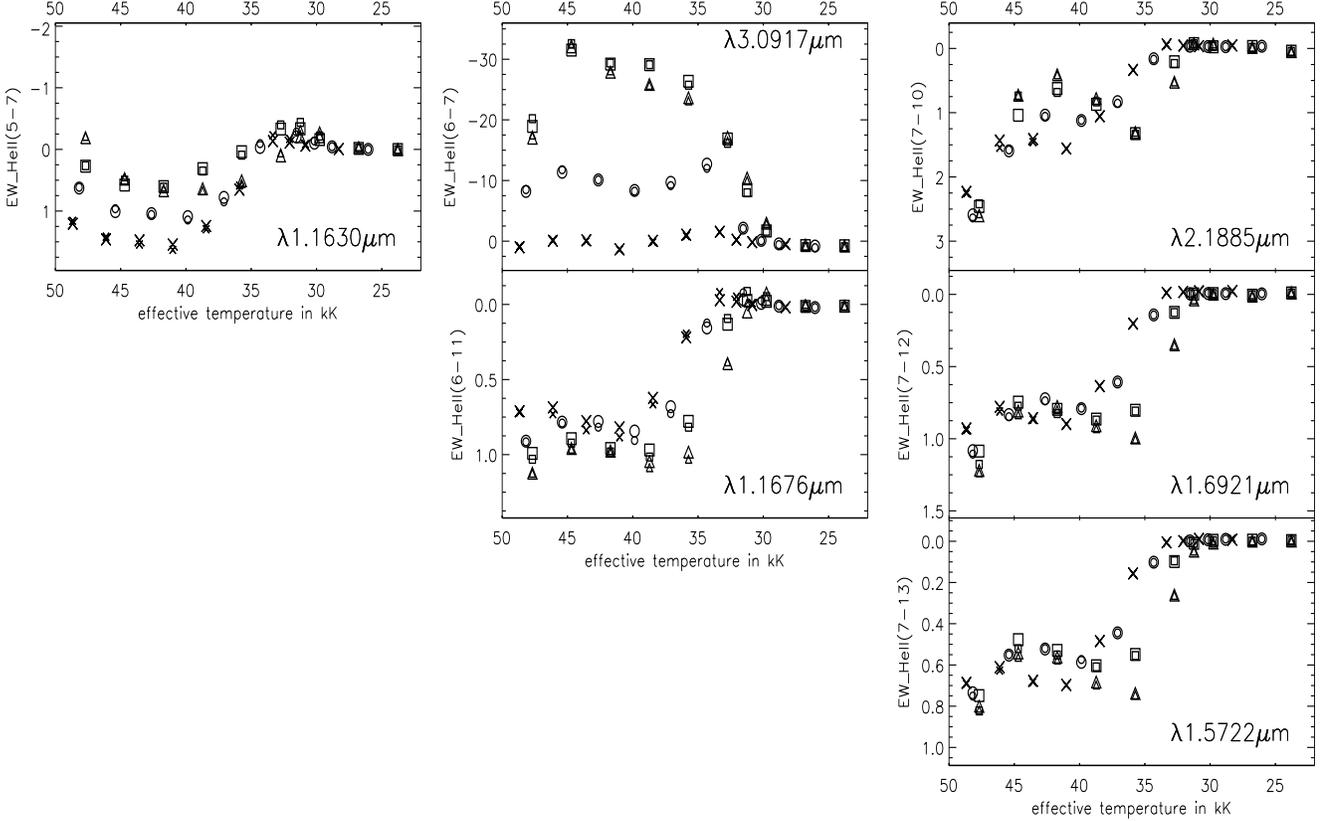}}}
    \caption{Equivalent width predictions (in \AA) for \heii\ lines are
    plotted for dwarfs (crosses), giants (circles), supergiants (squares)
    and lower surface gravity supergiants (triangles). Small symbols
    are for a turbulent velocity of 10 \kmsec, large symbols for
    turbulent velocity of 20 \kmsec.}
    \label{fig_heii}
    \end{center}  
    \end{figure*}

Several \heii\ lines are present in the different observational
bands. The strongest are the (5-7) and (6-11) transitions in the
J~band, (7-12) and (7-13) in the H~band, (7-10) in the K~band and
(6-7) and in the L~band. The behaviour of these lines can be split
into three regimes following the ionisation of \heii.  The lines first
appear at about 30\,kK and increase in absorption strength up to
$\sim$ 40\,kK. The exact temperature of this maximum absorption
depends on gravity, ranging from 41\,kK for dwarfs to 36\,kK for
supergiants. For temperatures $\teff \gtrsim$ 45\,kK, the lines weaken
again. All high-temperature models show \heii\ profiles in which the
line core is reverted in emission, as a result of a temperature
inversion in the line forming region. A temperature inversion is not
present in the highest \teff\ models. These therefore show normal
absorption profiles. This explains the increased absorption seen in
these hottest models. In giants and supergiants the strong $\alpha$
line (6-7) at \lam\,3.0917 \mum\ is formed in the stellar wind causing
a strong emission profile. For the hottest supergiant model the line
shows a decrease in emission that is the result of a lower wind
density (as discussed in Sect.\,\ref{sec_hyd}). The behaviour of
\heii\,\lam\,3.0917 is complex as it is blended with the (8-11) line,
which shows a singular behaviour as a result of it being interlocked
with the (6-7) transition. For those cases where the (6-7) line is in
emission, the (8-11) line is in absorption, while for an absorption
profile for (6-7) the blending line is in emission. All luminosity
classes suffer from this blending effect, however, for the supergiants
the (6-7) dominates the equivalent width in such a way that no dip in
EW occurs at $\sim$ 41\,kK. Wind emission is also important for the
(5-7) line.

\noindent The relatively weak \heii\ lines may serve as temperature
indicators, whereas the stronger lines are mostly sensitive to
luminosity class, similar to hydrogen lines. For temperatures between
33 and 42\,kK these lines are also sensitive to gravity.  Like for the
hydrogen lines, effects of micro turbulence are relatively modest.

\subsection{\hei\ lines}
\label{sec_hei}

    \begin{figure*}[!t]
    \begin{center} 
    \resizebox{\linewidth}{10.48cm}{\rotatebox{90}{\includegraphics{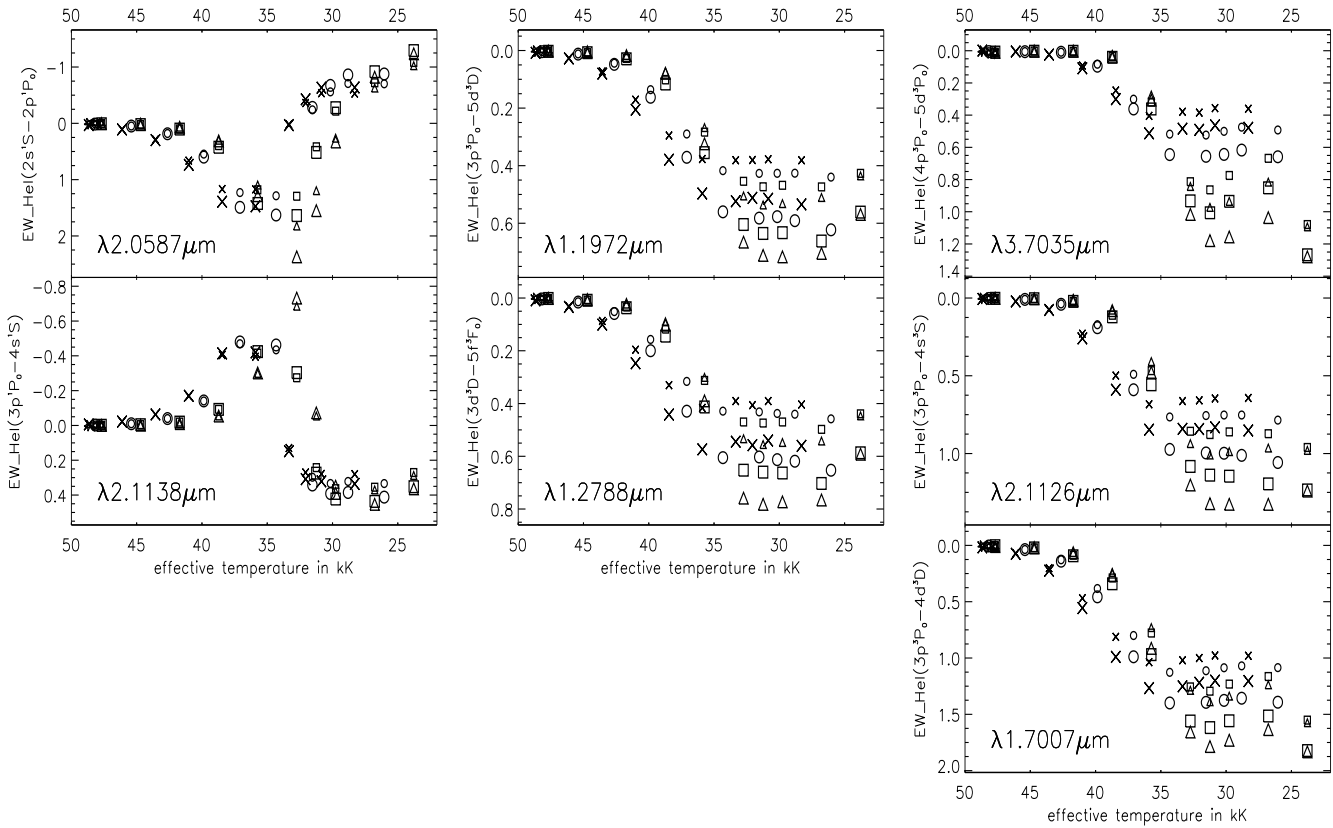}}}
    \caption{Equivalent width predictions (in \AA) for \hei\ are plotted for 
             dwarfs (crosses), giants (circles), supergiants (squares) 
             and lower surface gravity supergiants (triangles). Small 
             symbols are for a turbulent velocity of 10 \kmsec,
             large symbols for turbulent velocity of 20 \kmsec.}
    \label{fig_hei}
    \end{center}
    \end{figure*}

Numerous \hei\ lines are present in the near-infrared. Six of them
were observed in the J~band spectrum of a B0\,Ia star by Wallace et
al.\ (1997). We concentrate on the strongest lines present at
\lam\,1.1972 and 1.2788 \mum\ in the J~band, at \lam\,1.7007 in the
H~band, at \lam\,2.0587, 2.1126 and 2.1138 in the K~band, and at
\lam\,3.7036 \mum\ in the L~band. The line trends are plotted in
Fig.~\ref{fig_hei}.

We first discuss the triplet transitions. These show a normal
excitation and ionization behaviour, producing a broad peak in
absorption strength near $\sim$ 30\,kK. For higher temperatures the
lines weaken, and disappear somewhere between 40 and 50\,kK. Lines are
weaker in lower gravity models that have $\teff \gtrsim$ 35\,kK, a
situation that is reversed at lower temperatures. The lines at
\lam\,1.7007, 2.1126 and 3.7036 \mum\ are the most sensitive to
gravity. For a higher microturbulent velocity the equivalent width of
the \hei\ lines increases, as expected. We note that for the lines at
\lam\,1.1972, 1.2788 and 3.7035 \mum\ we do not have Stark broadening
tables available, implying we underestimate the strength of these
lines. The neglect of pressure broadening also explains why these
lines appear more affected by a change in turbulence.

    
The singlet lines at \lam\,2.0587 and 2.1138 \mum\ show a very
different behavior as a result of a strong coupling to the strength of
the ultraviolet line-blanketed continuum through the resonance
transitions at 584 \AA\ and 537 \AA\ (see Najarro et al. 1994). The
first transition reaches the upper level of the \lam\,2.0587 line, the
second one that of the lower level of the \lam\,2.1138 line. This
causes the ``inverted'' behaviour of these two lines as seen in
Fig.~\ref{fig_hei}. The weakening of both lines at $\teff
\gtrsim$~35\,kK is the result of progressive ionisation. The exact
location of this peak absorption strength (for \hei\,\lam\,2.0587) and
peak emission strength (for \hei\,\lam\,2.1138) depends somewhat on
gravity, but ranges from 38\,kK for dwarfs to 33\,kK for supergiants.


Most of the \hei\ lines discussed are expected to be useful
temperature diagnostics, especially at temperatures beyond
$\sim$~35~kK. At lower \teff\, the lines show a strong dependence on
gravity and/or turbulent velocity.

\section{Comparison with observations}
\label{sec_obs}

\begin{figure*}[!t]
    \begin{center}
    \resizebox{\linewidth}{14cm}{\rotatebox{90}{\includegraphics{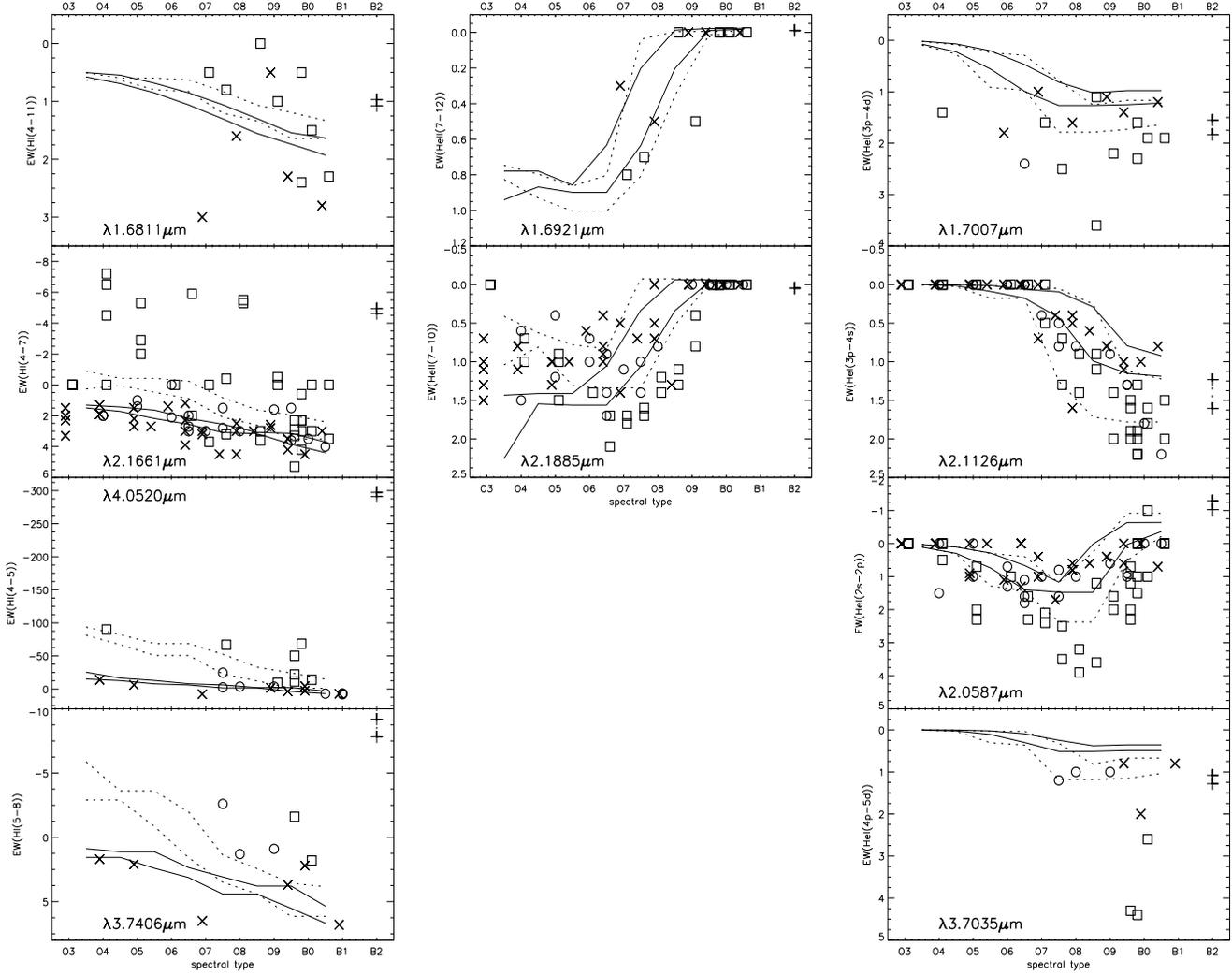}}}
    \caption{The maximum and minimum EW predictions (in \AA) are
     plotted for dwarf (solid lines), and supergiant (dotted lines)
     models. Models beyond the bi-stability jump are denoted by
     plusses. Observations from Hanson et al. (1996), Blum et
     al. (1997), Hanson et al. (1998), Zaal et al. (2001) and Lenorzer
     et al. (2002a) are plotted with crosses for dwarfs, circles for
     giants and squares for supergiants.  Typical error on the
     observations is $\pm$ 0.5 \AA.}
    \label{fig_obs}
    \end{center}
    \end{figure*}

Calibration of the trends predicted by the models requires a large
data set of near-infrared spectroscopic observations of O-type stars
with well known properties obtained from optical and ultraviolet
studies.
Observations of near-infrared lines are still scarce, although some
effort has been made in this direction over the last decade. In
particular, we make use of the K-band atlas of Hanson et al. (1996),
the H-band collection of spectra of Hanson et al. (1998) and the
ISO/SWS atlas of Lenorzer et al. (2002a) covering the
L-band. Additional measurements were gathered from Blum et al. (1997)
and Zaal et al. (2001) for the \hei\ line at \lam 1.7007 and the \bra\
line, respectively. This collection of observations allows a first
limited comparison with models. Observed equivalent widths are plotted
in Fig.~\ref{fig_obs} together with the model trends. These trends
give the minimum and maximum EW values for the dwarf (solid lines) and
supergiant models (dotted lines), assuming that the spectral types
attributed to the models are accurate to within one subtype.

Considering the hydrogen lines (left panel), the general trends are
fairly well reproduced by the models, though the observations of lines
affected by the stellar wind (\bra, \brg) show a much larger scatter
than is produced by the range of parameters that we have investigated.

\subsection{The \brg\ problem.}
The line for which a large observed scatter is clearly seen is
\brg. Note that the only model in the \brg\ panel that produces an
emission equivalent width that is comparable to the highest values
that are observed is the one for the coolest \teff\ (represented by a
plus sign). This model is at the cool side of the bi-stability jump
and has a much higher wind density.
One would expect a range of wind densities in O-type supergiants of
given spectral type, as these show a considerable spread in luminosity
(and therefore mass loss). The observations, however, lie
systematically above the prediction of typical Ia stars. This may
either suggest a strong underprediction of the wind density (which
seems unlikely on the basis of optical and ultraviolet studies) or
that clumping of the stellar wind is important for near-infrared wind
lines. A clumped medium would lead to an increase of the emission line
strength, as clumping results in an increased recombination rate
inside the clumps and in an enhanced optical depth
($\langle\rho^2\rangle\, >\, \langle\rho\rangle^2$). Although this
seems to be a plausible argument, \brg\ poses two additional
problems. Note at first that due to the lower oscillator strength and
the lower occupation numbers, \brg\ should form {\it inside} the \ha\
emitting volume (the gf-value of \brg\ lies in between the gf-values
of \ha\ and \hb). Actually, for almost all of our models it turned out
that \brg\ is formed in the same region as \hb. Since this region lies
close to the sonic one, the degree of clumping should be
small. Second, our synthetic \brg\ lines for stronger winds display
well developed P~Cygni type of profiles, whereas almost all
observations show a pure emission profile. Interestingly, both
problems have also been found in analogue investigations by Jokuthy
(2002) performed by means of {\sc fastwind}, an independent code that
uses similar physics as decribed here. Note particularly that optical
analyses of a large sample of O-stars using the latter code gave no
evidence of significant clumping effects for \hb\ (cf. Repolust et
al.\ 2004).

Thus, although clumping cannot be excluded, the degree of mismatch
between synthetic and observed profiles and EWs indicates that some
additional (physical) processes might be involved. Indeed, the
difference between observed and synthesized profile shapes points to a
line source function which should be closer to LTE than presently
calculated (P~Cygni shaped recombination lines can arise only due to
departures from LTE, whereas LTE results in pure emission profiles,
cf. Puls et al. 1996). Additionally, the sensitivity of the line
source function on small changes in departure coefficients (see
Eq.~\ref{sline_ir}) renders the possibility that the degree of
mismatch is not as large as indicated on a first glance, and that some
subtle modifications in the underlying physics might cure the problem.
To this end, an update of presently used hydrogen line-collision rates
as proposed by Butler \& Przybilla (in preparation for A\&A) might
help in driving the transition closer to LTE and increasing the line
emission.

Another evidence that the above problem is somewhat peculiar follows
from the fact that the \bra\ measurements appear to follow the
predictions fairly well. If clumping would play a role in the
formation of \brg, this would suggest that the clumping factor varies
throughout the wind, reaching a maximum in the lower wind regions and
decreasing again farther out. Since \bra\ orginates in a larger volume
of the wind, it would be relatively less affected by clumping.


\subsection{\heii\ lines}
Observations of the \heii\,(7-12) and (7-10) lines are compared with
predictions in the middle panel of Fig.~\ref{fig_obs}. For the (7-12)
line only few observations are available, limiting a meaningful
comparison. For those few available data points, it appears that the
models reproduce the observations reasonably well. A much better
comparison can be made for the (7-10) line at \lam\,2.1891
\mum. Though the observed global trends are recovered, we find that
for dwarf stars with spectral types earlier than O7 this line is
overpredicted by about 0.5 \AA, similar to the optical \heii\
line. For supergiants of spectral type later than O6, the (7-10) line
at \lam\,2.1891 is underpredicted by up to $\sim$ 1\,\AA\ for
supergiants of spectral type later than O6. The reason for this
behaviour is most likely linked to the uncertainty in the density
structure. We find that the line cores of the \heii\ lines are
typically formed at 10 to 100 \kmsec, i.e., at or even beyond the
sonic point. This implies that the dominant line contribution is
formed in the transition region from the photosphere to the wind, for
which we essentially do not have a self-consistent solution for the
density structure (see Sect.~\ref{sec:limitations}). The results
suggest that especially for supergiants the simple transition from an
exponentially decaying density to that implied by a $\beta$-type
velocity law adopted in our models is not correct (and not as much the
simplified hydrostatic density structure in the photosphere itself).

\subsection{\hei\ lines}
For the \hei\ lines a comparison of the \lam\,1.7007, \lam\,2.1126,
and \lam\,2.0587 lines is feasible (right panel). Though observational
data is also available for the \lam\,3.7035 \mum\ line, this data is
limited to only a few stars. Moreover, our models do not account for
the Stark broadening of this line. We therefore decided to exclude
this line from the comparision. For the remaining three lines both the
trend in spectral type and in luminosity class are reproduced by the
predictions. In particular, the \hei\,\lam 2.1126 line behaves
well. Note that this 3p-4s transition is plotted as the sum of the
\lam\,2.1126 and \lam\,2.1138 transitions, which are blended with each
other in the medium resolution spectra of Hanson et al. (1996). Also
the \lam\,2.0587 line behaves well, though most supergiants appear to
show a deeper absorption by up to $\sim$ 1 \AA. The same discrepancy
for supergiants seems present in the \lam\,1.7007 line. The cause of
the systematic differences in absorption strength are again most
likely connected to the density stratification in the transition from
photosphere to wind.

\noindent
\begin{center}
---
\end{center}

We conclude from this comparison that the models reproduce the global
trends of H
(with severe problems in \brg), \hei, and \heii\ lines, but that the
strength of helium lines tends to be underpredicted for the
supergiants with spectral type later than O6. The models that best
approach the He line strengths in Ia stars are the ones with low
gravity, i.e.\ not with canonical values. This does not imply that
these stars have a much lower mass than expected, but shows that the
density structure in the transition region and lower wind is better
represented by a low gravity in Eq.~\ref{eq:vlaw}.
 We also like to remind the reader that all models have been
calculated with normal Helium abundance, which introduces additional
uncertainties in those cases when the stellar atmosphere contains
enhanced He, e.g.\ due to rotational mixing.


\subsection{Lines of carbon and nitrogen}

The majority of the lines present in the near-infrared spectra of hot
stars are produced by hydrogen and helium. In the K~band, a few strong
emission lines are attributed to carbon and nitrogen ions. A triplet
of lines around 2.08 \mum, identified as the \civ\ (3p-3d)
transitions, appear strongly in emission in the spectra of stars with
spectral type O4 to O6.5. In our models, however, these lines are
strongly in absorption at spectral types earlier than O5.5. They
revert in emission only for later types and disappear at O9. It is
obvious that our models fail in reproducing these transitions.
Matching of the carbon lines is known to be problematic in O-type
stars (e.g. Lamers et al.\ 1999) as a result of uncertainties in the
ionization structure of this element. The carbon ionization is
sensitive to the amount of line blanketing in the extreme ultraviolet
(EUV) part of the spectrum, notably in the ionizing continua of \ciii\
and \civ. However, the 2.08 \mum\ \civ\ lines are observed to be
narrow with a FWHM of about 43 \kmsec\ (Bik et al.\ in prep.)
indicating that they originate in the photosphere and are not filled
in by a contribution from the wind. A change in the abundance of \civ\
due to metallicity would influence the strength of the 2.08 \mum\
lines without reverting them into emission profiles.  The formation
region of the near-infared \civ\ lines largely overlaps with that of
the \hei\ line at \lam\,2.0587\,\mum. Our models qualitatively
reproduce the behaviour of the \hei\ line indicating that the local
parameters in this region are most likely reasonable. The levels
taking part in the \civ\ lines observed around 2.08 \mum\ are
populated through transitions located in the EUV. The inclusion of
\feiii, \ciii\ and \oiii\ had the direct consequence that the \civ\
lines reverted into emission in models with effective temperature from
about 40\,000 to 32\,000 K.
Though we account for approximately 30\,000 lines (see
Sect.~\ref{sec_grid}), this does likely not yet represent all of the
blanketing at EUV wavelengths. A more complete inclusion of EUV lines
is curently under investigation and may further reduce the \civ\
discrepancy at temperatures higher than 40\,000~K.

Our model atoms do not include data for \niii\,(7-8), expected to be
partly responsible for the observed emission near \lam\,2.1155 \mum.
Still, our predictions show an emission due to \ciii\,(7-8) at
\lam\,2.1151 \mum. The EW of this line peaks at 0.8 \AA\ for spectral
type O6.5. This is much weaker (1 to 5 \AA) than observed, though it
may explain some observations for spectral types O6 and later.

\section{Diagnostics}
\label{sec_param}

In the Morgan \& Keenan (MK, see Morgan \& Keenan 1973) classification
scheme stars of spectral type O are defined by the presence of \heii\
lines. Subtypes are defined on the basis of the relative strength of
the \hei\ and \heii\ lines. The luminosity class of O-type stars is
based on optical \siiv\ and \hei\ lines (e.g. Walborn 1990). A
quantification of the spectral type calibration was proposed by Conti
\& Alschuler (1971), Bisiacchi et al. (1982) and Mathys (1988). In
this section, we investigate the correlation between optical and
near-infrared line ratios. This may lead to an extension of the
quantitative spectral classification to near-infrared lines.

The advantage of the near-infrared spectral range over the optical is
that it contains a range of hydrogen lines from different series and
from high levels in the atom. Although this implies an additional
diagnostic potential (see Lenorzer et al. 2002b), it does require
spectra of sufficient quality, both in terms of signal-to-noise and
spectral resolution, in order to capitalize on this. Also, $\alpha$,
$\beta$, and $\gamma$ lines are stonger than their counterparts in the
Balmer series in the optical.  This may turn out to be beneficial,
especially for wind density
determinations based on \bra\ (see Sect.~\ref{sec_mass}). The helium
lines in the optical and near-IR are about equally strong, so in
principle they may serve equally well for spectral classification. In
practice the data quality is usually somewhat less in the near-IR as
present day detectors for this wavelength regime have a poorer quantum
efficiency than optical instruments do.

\subsection{Spectral Type}
\label{sec_st}

\begin{figure}[!t]
    \resizebox {\linewidth}{14cm}{\includegraphics{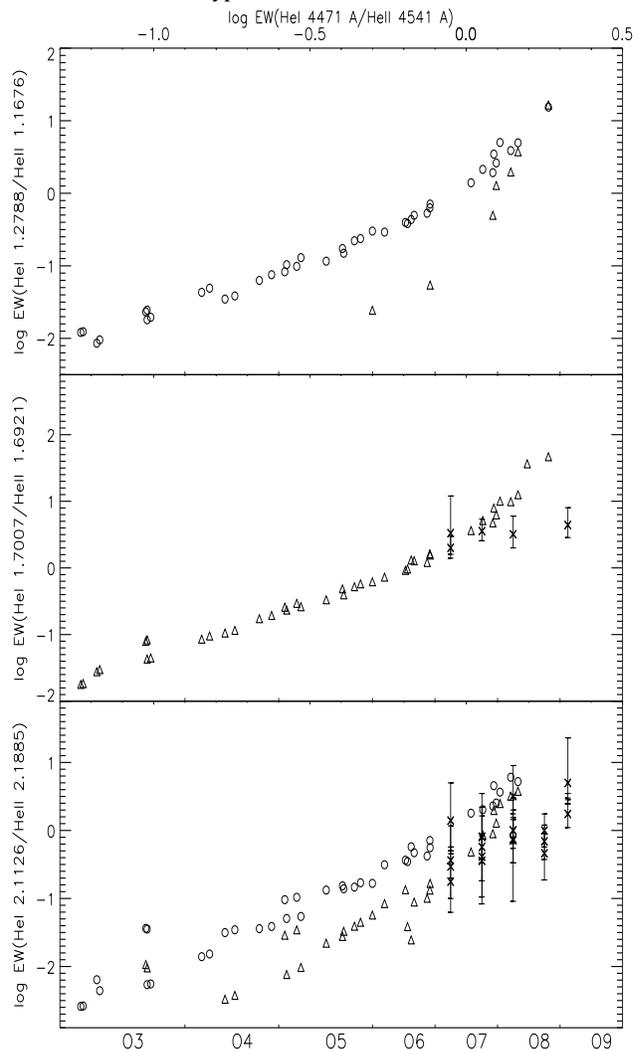}}
     \caption{The correlation between optical and near-infrared
              \hei/\heii\ equivalent width ratios which may be used
              to calibrate spectral types. Circles denote EW 
              predictions for a relatively high spectral
              resolution ($R \sim 4\,500 - 6\,000$); triangles give
              predictions for medium resolution ($R \sim 1\,000$).
              The plus symbols give observed ratios.
             }
    \label{fig_st}
    \end{figure}

When correlating the optical and near-infrared line behaviour, we
first notice that O-type stars are no longer {\em defined} by the
presence of \heii\ lines if one concentrates on only the near-infrared
window.  Indeed, at these wavelengths, most \heii\ lines disappear
around spectral type O8.5. Consequently, near-infrared \hei/\heii\
line ratios can only be measured for earlier spectral types. This is
to be expected as the near-infrared continuum is formed further out in
the atmosphere of O-type stars, where the temperature is lower.

Predicted \hei/\heii\ equivalent width ratios are presented in
Fig.~\ref{fig_st} for lines in the J, H and K band. In the L-band no
suitable helium lines were found. No distinction between the
luminosity classes was made and values applicable for both high- and
medium-resolution spectra are plotted. High spectral resolution
(circles in Fig.~\ref{fig_st}) implies $R \sim 6\,500$ and 4\,500 for
the J and K band. With medium resolution (triangles) $R \sim 1\,000$
is implied. In the J-band (top panel) we find that the ratio
\hei~\lam 1.2788/\heii~\lam 1.1676 may serve to determine the
spectral type. This only works at high resolution, as else the
\hei\,1.2788 line starts to blend with \ciii\,1.2794, rendering this
diagnostic unusable. Note that for \hei\,\lam 1.2788 no Stark
broadening is taken into account. This will affect the slope of the
relation and at present prevents a reliable use of this ratio as a
temperature diagnostic.

In the H band (middle panel) the best candidate line ratio for
determining spectral type is \hei\,\lam 1.7007 / \heii\,\lam 1.6921.
In practice, the optical ratio \hei\,\lam 4471 over \heii\,\lam 4542
is found to be of practical use for ratios inbetween about $-1$ and
$+1$ dex; otherwise one of the two lines gets too weak. The same
applies to the identified H band ratio, implying that it can be used
to determine the spectral type of O4 through O8 stars.

Overplotted in Fig.~\ref{fig_st} are observed \hei / \heii\ ratios
from Hanson et al. (1998) in the H band and from Hanson et al. (1996)
in the K band. In the H band only few observations are available for
spectral type O7 through O9. These seem to form an extension of the
predicted slope that is found for stars of types O7 and
earlier. However, at spectral type O8 the predicted curve turns
upward, i.e. the models appear to overpredict the helium line
ratio. In the K band, more data is available, although also only for
types O7 and later as otherwise the \hei\,\lam 2.1126 line is too weak
to be measured. The available data is mainly for supergiants. We find
that the observations extend the predictions of the low surface
gravity models for stars of spectral type O7 or earlier, and
underpredict the ionisation for cooler stars. On the basis of such a
limited comparison it is difficult to draw firm conclusions. We again
note that the observed ratios appear in better accordance with the low
gravity models. This likely implies that for supergiants the density
structure in the transition region and lower part of the wind connects
more gradual (i.e. with a smaller density gradient) than is assumed in
our standard supergiant grid, rather than that it implies an
overestimate of the mass.

The predicted near-infrared equivalent width ratios presented in
Fig.~\ref{fig_st} correlate well with the optical ratio and show a
steeper dependence on spectral type than does \hei\,\lam 4471 over
\heii\,\lam 4541.
These near-infrared line ratio may serve to determine the spectral
type. As for our model predictions, we note that the predictions for
\hei\, \lam 1.2788 do not account for Stark broadening
effects. Therefore, the J band ratio is less steep than presented,
yielding it will yield spectral types that are systematically too
early. The H-band and K-band ratio are also to be taken with care as
our models do not perfectly reproduce the observed line strengths (see
Sect.~\ref{sec_obs}). Still they give a reasonably good idea of the
observational requirements needed to derive quantitative information
on the spectral type of hot stars from near-infrared spectroscopy
alone.
We conclude that a derivation of the spectral sub-type of O stars from
near-infrared helium line ratios is in principle feasible for good
quality spectra in the J, H, and K band and for stars that have
spectral types in the range O4 to O8.

\subsection{Surface gravity}
\label{sec_lumi}

Several of the strongest lines in the near-infrared, such as \bra,
\brg, \pfg, and \heii\,\lam 3.0917, appear to be very sensitive to the
surface gravity. This is, however, a consequence of the relation we
assume between gravity and mass-loss rate. These lines are mainly
formed in the wind and may serve as an indicator of the wind density
(see next subsection).

The lines showing a direct dependence on surface gravity are the
weaker lines, formed near the stellar photosphere. Most \hei\ and a
few \heii\ lines are found to be sensitive to this parameter. The
\heii\,(6-11), (7-10), (7-11) and (7-12) and the \hei\,(2s-2p) lines
are stronger in absorption for lower gravity at temperatures between
30 and 40 kK (see Sect.~\ref{sec_trend}). The same holds for most
\hei\ lines between 25 and 35 kK. Unfortunately, no well defined
monotonic correlation could be extracted.

\subsection{Wind density}
\label{sec_mass}

Mass-loss rates of O-type stars can be determined from ultraviolet
resonance and subordinate lines, \ha, and radio flux measurements.
For a recent review on these methods, including a discussion of their
individual pros and cons, see Kudritzki \& Puls (2000). A relatively
simple method to derive \mdot\ is to use the net equivalent width of
\ha\ (see e.g. Klein \& Castor 1978, Leitherer 1988,
Puls et al. 1996) 
and to correlate it with the equivalent width invariant
\begin{equation}
   Q = \frac{\mdot}{R^{3/2} \teff^{2} \vinf}~,
\end{equation}
first introduced by Schmutz, Hamann \& Wessolowski (1989) for a fixed
temperature, and extended to include \teff\ by Puls et al. (1996) and
de Koter, Heap \& Hubeny (1998). The invariant essentially expresses
that \ha\ is formed by the recombination mechanism, therefore its
strength will be approximately proportional to the mean column mass
$\overline{\rho}^{2} \Rstar \sim \mdot^{2} / (\Rstar^{3} \vinf^{2})$.
In principle, the same strategy can be applied using the \bra\
line. Relations between the EW of \bra\ and the mass-loss rate have
already been proposed, based on model predictions (Schaerer et
al. 1996); and on observations (Lenorzer et al. 2002a). This line is
intrinsically stronger than \ha, which means that the photospheric
absorption has a smaller impact on its EW. We have opted not to
correct for a photospheric contribution, as a ``true'' photospheric
component of this line can only be defined if a core-halo
approximation is adopted (i.e. a separate treatment of the stellar
photosphere and wind), which is not physically realistic in the
near-infrared regime. Recall that most near-infrared lines are formed
mainly in the transition region.

\begin{figure}[!t]
 \begin{center}
    \resizebox {\linewidth}{7.5cm} {\rotatebox{90}{\includegraphics{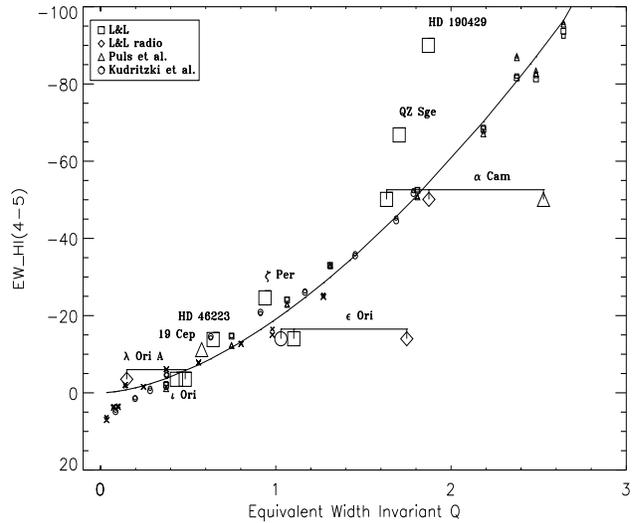}}}
    \caption[fig_mass]{The predicted equivalent widths for dwarfs
    (denoted by crosses), giants (circles) and supergiants (squares)
    and lower surface gravity supergiants (triangles) are overplotted
    with observations from Lenorzer et al. (2002a), for which Q was
    derived from Lamers et al. (1999), Puls et al. (1996), and
    Kudritzki et al. (1999). $Q$ is given in units of 10$^{-20}$
    \msunyr\ per \rsun$^{3/2}$ per K$^{2}$ per \kmsec.}
    \label{fig_mass}
    \end{center}
\end{figure}

The correlation between predicted \bra\ equivalent width and $Q$ is
given in Fig.~\ref{fig_mass}. The units of the constituents of the
latter quantity are \mdot\ in \msunyr, stellar radius in \rsun, \teff\
in Kelvin, and \vinf\ in \kmsec.  The measurements are from 4.0 to 4.1
\mum\ and include a number of blends from weak \hei\ and \heii\
lines. This causes most of the modest scatter in the model results;
overall the correlation is very good. A best fit (overplotted using a
solid line) to all negative EW values yields
\begin{equation}
   \log {\rm EW}\,({\rm Br}\alpha) = (32.4 \pm 0.5) + 
      (1.55 \pm 0.03) \log Q~.
\end{equation}
The divergence from the fit of models with mass-loss rates below about
$10^{-7}$ \msunyr\ marks the transition to profiles dominated by
photospheric absorption. In the wind dominated regime the fit function
recovers the $Q$ value to within 0.05 to 0.15 dex for $0.5 <
Q/10^{-20} < 1.0$, and to within 0.04 dex for larger values.
The reader may note that the above correlation has been evaluated for
the models as described in Sect.~\ref{sec_grid}, i.e. for a wind
velocity law with $\beta = 0.8$ and $1.0$ for dwarfs and supergiants,
respectively. Since \bra\ is strongly density dependent, different
values for $\beta$ will introduce certain deviations from this
relation (cf.\ Puls et al.\ 1996).

In Fig.~\ref{fig_mass} we overplot the observed equivalent widths of
nine giants and supergiants. The $Q$ values corresponding to these EW
are from Puls et al. (1996), Lamers et al. (1999) and Kudritzki et
al. (1999) (see figure caption for details). The bars connect
mass-loss determinations for the same stars by different authors,
sometimes using different methods (\ha\ or radio). The bars therefore
give an indication of the uncertainties involved in deriving
\mdot. Typical errors on individual measurements are about 0.3 dex in
the $Q$ parameter. In seven out of nine cases the determinatios are in
good agreement with the trend. This ignores the outlying \mdot\
results for $\alpha$\,Cam and $\epsilon$\,Ori. The observed EW for the
two Iaf supergiants in the sample, HD\,190429A and QZ\,Sge, are
clearly above the trend. These two stars show \heii\ emission lines
that are much stronger than the ones produced in our supergiant
models. One of these \heii\ lines is included in the measurement of
the \bra\ equivalent width, explaining the difference.
 

\section{Conclusions}
\label{sec_conclu}

We have confronted observations of near-infrared spectra with
predictions using the current state-of-the-art model atmosphere code
{\sc cmfgen} of Hillier \& Miller (1998). This study was prompted by
the fact that an increasing number of spectra of embedded O-type stars
is presently becoming available, allowing the study of stars hidden
beyond tens of magnitudes of circumstellar or foreground visual
extinction. Though first attempts to make an inventory of the
potential and predictive power of models in this wavelength range have
been undertaken (Najarro et al. 1999), a systematic approach using a
large grid of models was so far missing.  We summarize the main
conclusions of this study.

\begin{enumerate}

\item The general trends and strengths of the lines are fairly
      compatible with observed properties. Near-infrared lines are
      typically formed in the transition region from photosphere to
      stellar wind, or even in the lower part of the wind (up to a few
      times the sonic velocity). The largest discrepancies are found
      for helium lines in supergiant spectra, which are systematically
      underpredicted. As low gravity models appear to reproduce these
      lines much better, it is expected that for these stars the
      connection of the photosphere to the wind is much smoother,
      i.e. the density changes more gradually, than is adopted in our
      models. (It does not imply that masses for supergiants are
      systematically underestimated).

\item The models show a good correlation between optical and
      near-infrared helium line ratios in the J, H and K band.
      Notably, the 
      \hei~\lam~1.2788 / \heii~\lam~1.1676,
      \hei~\lam~1.7007 / \heii~\lam~1.6921, and
      \hei~\lam~2.1136 / \heii~\lam~2.1885 \mum\ ratios
      appear good candidates. Application of these
      near-infrared ratios requires good quality data and are only
      applicable for stars with spectral type O4 to O8. Cooler stars
      do not show \heii\ lines. This is to be expected as the
      near-infrared continuum is formed further out in the atmosphere,
      relative to where the optical spectrum originates.

\item The \bra, \brg, and \heii\,\lam 3.0917 lines in giants and
      supergiants are predominantly formed in the stellar
      wind. Comparison with observations shows that the \brg\ lines
      are systematically underpredicted, while in most cases \bra\ is
      reproduced well.
      We have argued that \brg\ might be affected by an
      inhomogeneous density distribution or ``clumping'', while the
      stronger \bra\ line suffers much less from this effect, which
      would suggests a distance dependent clumping.
      Additional arguments concerning line formation depth and
      profile shape of \brg, however, point also to presently 
      unknown processes which have to be identified in
      future investigations.

\item We find that the \bra\ line is a fairly good diagnostic for the
      wind density and we provide a means to correlate the measured
      equivalent width of this line to the wind density invariant $Q =
      \mdot / R^{3/2} \vinf \teff^{2}$. The relation, however, depends
      on the \heii\ abundance in the wind, as the \bra\ line is
      contaminated by a potentially strong \heii\ line and may
      strongly affect mass loss predictions for Iaf stars.

\end{enumerate}

\acknowledgements{The authors are very grateful to D.J.~Hillier who
courteously provided his {\sc cmfgen} code, as well as help in setting
up the grid. We would like to thank F.~Najarro for inspiring and
constructive discussions and D.~Schaerer whose careful reading and
suggestions helped improving the quality of this paper. We also wish
to thank the referee A.~Herrero for his critical reading and
constructive comments.}

{}

\end{document}